\documentclass[twocolumn,prl]{revtex4}
\usepackage{stmaryrd}
\usepackage{amsmath}
\usepackage{amssymb}
\usepackage{graphicx}
\usepackage{dcolumn}
\usepackage{bm}
\usepackage{multirow}
\usepackage{array}

\begin{document}

\begin{titlepage}

\title{Superconducting Graphene: the conspiracy of doping and strain}

\author{Chen Si$^{1,2}$, Zheng Liu\footnote{This author contributed equally as the first author.} ,$^2$ Wenhui Duan,$^1$ and Feng Liu$^2$\footnote{fliu@eng.utah.edu}}
\address{$^1$Department of Physics and State Key Laboratory of Low-Dimensional Quantum Physics, Tsinghua University, Beijing 100084, People's Republic of China \\
$^2$Department of Materials Science and Engineering, University of Utah, Salt Lake City, Utah 84112, USA}

\date{\today}

\begin{abstract}
Graphene has exhibited a wealth of fascinating properties, but is also known not to be a superconductor. Remarkably, we show that graphene can be made a conventional Bardeen-Cooper-Schrieffer superconductor by the combined effect of charge doping and tensile strain. While the effect of doping is obvious to enlarge Fermi surface, the effect of strain is profound to greatly increase the electron-phonon coupling. At the experimental accessible doping ($\sim4\times10^{14}$ cm$^{-2}$) and strain ($\sim16$\%) levels, the superconducting critical temperature $T_{c}$ reaches as high as $\sim30$ K, the highest for a single-element material above the liquid hydrogen temperature. This significantly makes graphene a commercially viable superconductor.
\end{abstract}

\pacs{73.22.Pr, 74.10.+v, 74.20.Fg}


\maketitle

\draft

\vspace{2mm}

\end{titlepage}
Since its first making\cite{novoselov2004electric}, graphene has fascinated the scientific community by a seemingly endless discovery of extraordinary properties, such as electronically the highest carrier mobility with massless Dirac Fermions\cite{novoselov2005two, morozov2008giant}, optically the largest adsorption per atomic layer in the visible range\cite{bonaccorso2010graphene}, and mechanically the strongest 2D material in nature\cite{lee2008measurement}. However, graphene is considered not to be a good superconductor. In particular, two fundamental conditions of intrinsic graphene define an overall very weak electron-phonon coupling (EPC), rendering itself not to be a Bardeen-Cooper-Schrieffer (BCS) superconductor\cite{bardeen1957microscopic, bardeen1957theory}. First, graphene has a point-like Fermi surface (Dirac point) with vanishing density of states (DOS); second, it has a weak electron-phonon (e-ph) pairing potential. While the first condition can be obviously modified by doping; the second condition is not known amenable to change.

Several ideas related to doping have been proposed to induce superconductivity in graphene. For example, adsorption of alkali metal atoms on graphene has been found to introduce large DOS around the Fermi level as well as increase the e-ph pairing potential, and hence to enhance the e-ph coupling (EPC) for BCS superconductivity\cite{profeta2012phonon}. However, the EPC enhancement is largely due to the DOS and phonon modes of metal atoms, rather than the intrinsic properties of graphene. A recent theoretic work suggests that in a highly-doped graphene up to the point of Van Hove singularity the greatly increased e-e interaction can induce a pairing potential in the d-wave channel, possibly giving rise to chiral superconductivity\cite{nandkishore2012chiral}, but the critical superconducting transition temperature ($T_{c}$) is unknown and likely to be low.

In this Letter, we demonstrate, using first-principles calculations, that in combination with doping of either electrons or holes, biaxial tensile strain can greatly enhance the EPC of graphene in a nonlinear fashion so as to convert it into a BCS superconductor. Most remarkably, within the experimental accessible doping\cite{efetov2010controlling} and strain\cite{lee2008measurement} levels, $T_{c}$ can reach $\sim30$ K, the highest known for a single-element superconductor.

According to BCS theory\cite{bardeen1957microscopic, bardeen1957theory}, when the phonon-mediated attraction is strong enough to overcome the Coulomb repulsion, electrons form "Cooper pairs", leading to the emergence of superconductivity below $T_{c}$. The former is characterized by a dimensionless parameter $\lambda = N_{F}V_{ep}$, where $N_{F}$ is the electron DOS and $V_{ep}$ is the mean e-ph pairing potential at the Fermi level; the latter by a dimensionless parameter $\mu = N_{F}V_{ee}$, where $V_{ee}$ is the mean e-e repulsive potential at the Fermi level. Superconductivity occurs for $\lambda \gg \mu$, and $T_{c}$ increases with the increasing Fermi surface (more Cooper pairs be formed) and e-ph paring potential (easier Cooper pairs be formed). Because $\mu$ is rather material insensitive (see discussion below), generally, materials are classified into three regimes of EPC: weak $\lambda \ll 1$, intermediate $\lambda \sim 1$, and strong $\lambda > 1$; a good BCS superconductor requires $\lambda \geq 1$.

As a promising material for the next-generation electronic devices, the EPC in graphene has been extensively studied both theoretically and experimentally\cite{ferrari2007raman}. Because of a diminishing Fermi surface (a point for intrinsic graphene) and a very weak e-ph pairing potential (because of its high Fermi temperature of the massless carriers and high energy of the optical phonons), the EPC in graphene is found to be very weak. This feature is actually responsible for some of its other extraordinary properties like extremely high electrical\cite{chen2008intrinsic} and thermal conductivity\cite{balandin2008superior}. But on the other hand, it prevents graphene from being a BCS superconductor. To increase EPC of graphene ($\lambda$), one must increase the DOS ($N_{F}$) and/or the e-ph pairing potential ($V_{ep}$) at the Fermi level. Obviously, $N_{F}$ can be increased by doping of either electrons or holes. Figure 1a shows our calculated $N_{F}$ and $V_{ep}$ as a function of hole concentration ($n$) for a p-type graphene \cite{SI}. Both $V_{ep}$ and $\lambda$ increase with the increase of doping, and $\lambda \approx 0.19$ at a doping level of $6.2 \times 10^{14}$ cm$^{-2}$ (corresponding to doping of $\sim$1/3 of a hole per unit cell).

\begin{figure}[tbp]
\includegraphics[width=0.5\textwidth]{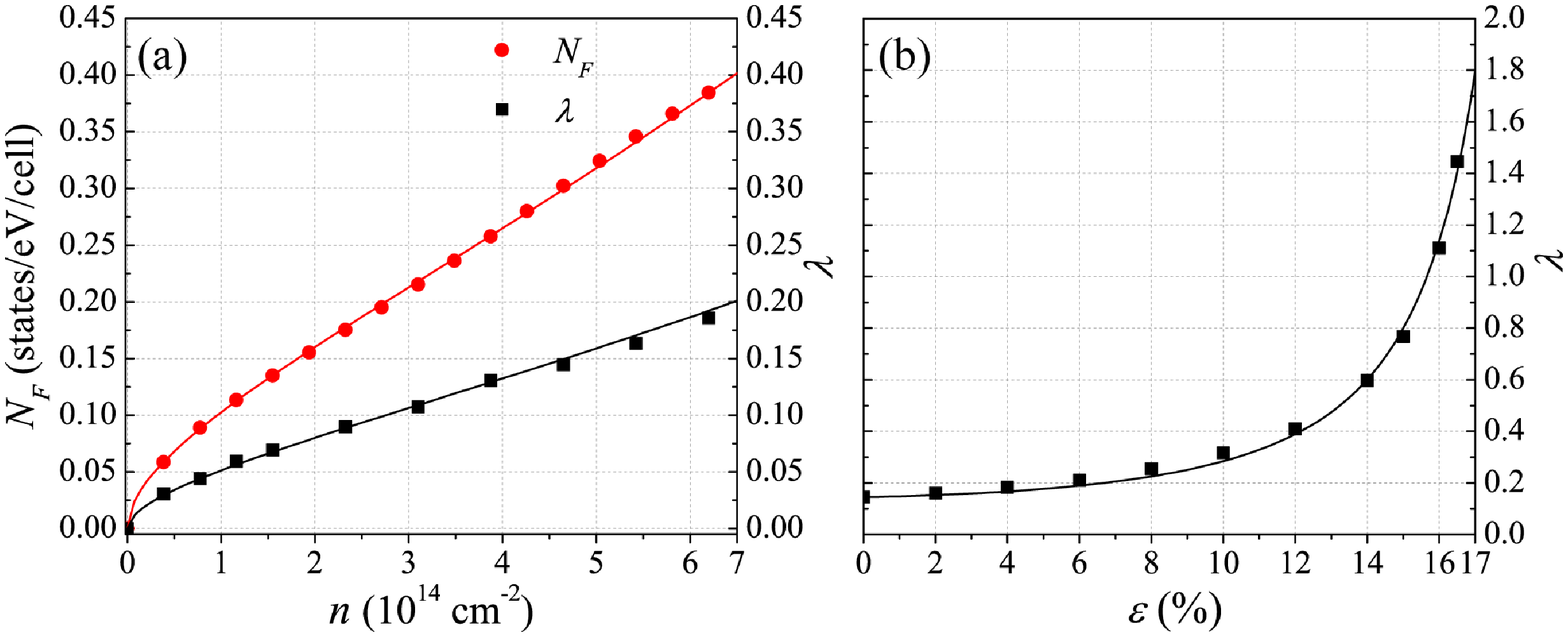}
\caption{\label{fig:fig1}  (Color online) (a) $N_{F}$ and $\lambda$ of p-type graphene under different doping level ($n$) calculated from first-principles. (b) $\lambda(\varepsilon)$ for the $4.65\times10^{14}$ cm$^{-2}$ hole-doped graphene. The solid lines are fits to the data. }
\end{figure}

Because we are interested in doping levels beyond the linear-Dirac-band regime, we expand $E(k)$ around the Dirac point to the second order\cite{neto2009electronic, wang2010manipulation}, $E = \alpha k+\beta k^{2}$, then we have $N_{F}=\sqrt{n}(a+b\sqrt{n})^{-1}$,	which gives a very good fit to the calculated $N_{F}$ with $a=11.93$ and $b=-2.19$, as shown in Fig. 1(a). Since $\lambda= N_{F}V_{ep}$, we found that $V_{ep}$ remains a constant, $\sim$0.5 eV, independent of doping, as shown by the very good fitting of $\lambda=0.5N_{F}$ shown in Fig. 1(a). Thus, we obtain a relation of   $\lambda(n)$ for the hole-doped graphene as
\begin{equation}
\lambda = 0.5\sqrt{n}(11.93-2.19\sqrt{n})^{-1} \label{eq:one}
\end{equation}
We noticed that the value of $V_{ep}\sim0.5$ eV is much smaller than the typical values found in BCS superconductors, such as 1.4 eV\cite{savini2010first, an2001superconductivity} for MgB$_{2}$ and 3 eV\cite{savini2010first, giustino2007electron} for B-doped diamond. This means that doping is a necessary but insufficient condition to make graphene a BCS superconductor.

The above results indicate that in order to make graphene a superconductor, one must find a way to increase $V_{ep}$ in addition to doping. It has been shown recently that applying biaxial tensile strain can significantly soften the in-plane optical modes of graphene\cite{marianetti2010failure}, hinting that it may also enhance $V_{ep}$ and hence the EPC. To explore this possibility, we have calculated $\lambda$ as a function of biaxial tensile strain ($\varepsilon$) for a hole-doped graphene at a $4.65\times10^{14}$ cm$^{-2}$ doping level, as shown in Fig. 1(b). Clearly we see $\lambda$ increases dramatically with the strain. In particular, at the 16.5\% of strain, $\lambda$ reaches as high as 1.45, entering the strong coupling regime, with a corresponding value of $V_{ep}\sim3.25$ eV, even larger than that in the B-doped diamond\cite{savini2010first, giustino2007electron}.

To understand such remarkable strain induced enhancement of EPC, we first recall that McMillan has shown that\cite{mcmillan1968transition} $\lambda \propto (\langle\omega^{2}\rangle)^{-1}$, where $\omega$ is the frequency of all the phonon modes contributing to the EPC. Interestingly, we found that for graphene there exists a characteristic phonon mode ($\omega_{0}$) that dominates the EPC. In general, the frequency of this characteristic mode must change with strain as $\omega_{0}(\varepsilon)=\omega_{0}(\varepsilon=0)+p\varepsilon+q\varepsilon^{2}$, where $p$ and $q$ are the first- and second-order phonon deformation potential, respectively; the second-order nonlinear term is needed here because of the large strain involved. Then, we can fit the $\lambda(\varepsilon)$ curve using an empirical formula
\begin{equation}
\lambda(\varepsilon) =0.5(1+t_{1}\varepsilon+t_{2}\varepsilon^{2})^{-2} \label{eq:two}
\end{equation}
and a very good fit is obtained with $t_{1}=-0.007$, $t_{2}=-0.002$, as shown in Fig. 1(b).

\begin{figure}[tbp]
\includegraphics[width=0.45\textwidth]{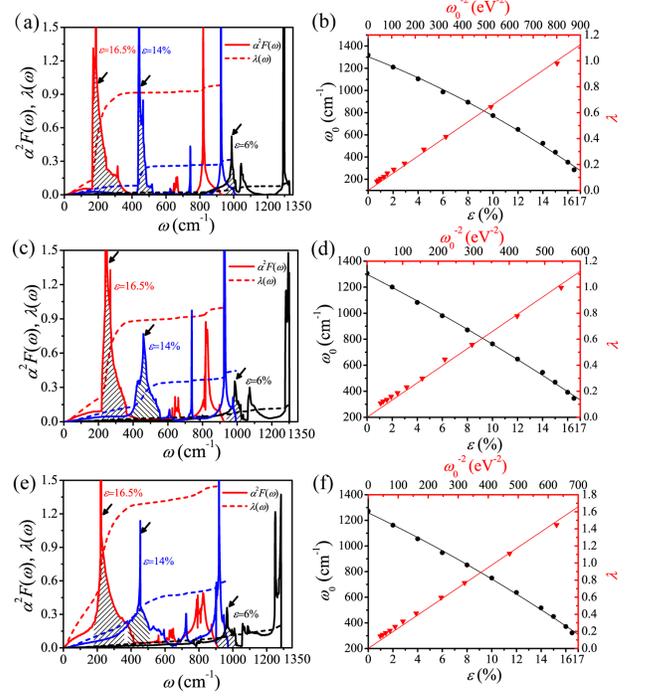}
\caption{\label{fig:fig2} (Color online) The Eliashberg spectral functions and $\omega_{0}$ as a function of strain in the hole-doped graphene. (a) $n=1.55\times10^{14}$ cm$^{-2}$ (c) $n = 3.10\times10^{14}$ cm$^{-2}$ (e) $n = 4.65\times10^{14}$ cm$^{-2}$ under 6\% (black line), 14\% (blue line) and 16.5\% (red line) strain. (b), (d) and (f), $\omega_{0}$ versus $\varepsilon$ (black dots) and ${\omega_{0}}^{-2}$ versus $\varepsilon$ (red triangle) for $n=1.55\times10^{14}$ cm$^{-2}$, $3.10\times10^{14}$ cm$^{-2}$ and $4.65\times10^{14}$ cm$^{-2}$. The solid lines are empirical fit to the data. }
\end{figure}

Next we perform a more vigorous analysis of the strain dependence of EPC in graphene. Figure 2 shows the Eliashberg spectral function $\alpha^{2}F(\omega)$, which describes the averaged coupling strength between the electrons of Fermi energy ($E_{F}$) and the phonons of energy $\omega$:
\begin{widetext}
\begin{equation}
\alpha^{2}F(\omega)= \frac{1}{N_{F}N_{k}N_{q}}\sum_{mn}\sum_{\textbf{q}\nu}\delta(\omega-\omega_{\textbf{q}\nu})\sum_{\textbf{k}}
|g_{\textbf{k}+\textbf{q},\textbf{k}}^{\textbf{q}\nu,mn}|^{2}\delta(E_{\textbf{k}+
\textbf{q},m}-E_{F})\delta(E_{\textbf{k},n}-E_{F}) \label{eq:three}
\end{equation}
\end{widetext}
and the frequency-dependent EPC function\cite{grimvall1981electron}
\begin{equation}
\lambda(\omega)=2\int_{0}^{\omega}\frac{\alpha^{2}F(\omega^{'})}{\omega^{'}}d\omega^{'} \label{eq:four}
\end{equation}
where the phonon frequency $\omega$ is indexed with wavevector ($\textbf{q}$) and mode number ($\nu$), and the electron eigenvalue $E$ is indexed with wavevector ($\textbf{k}$) and the band index ($m$ and $n$), and $g_{\textbf{k}+\textbf{q},\textbf{k}}^{\textbf{q}\nu,mn}$ represents the electron-phonon matrix element. For $1.55\times10^{14}$ cm$^{-2}$ (Fig. 2(a)), $3.10\times10^{14}$ cm$^{-2}$ (Fig. 2(c)) and 4.65 $\times10^{14}$ cm$^{-2}$ (Fig. 3(e)) hole doped graphene, the Eliashberg function is found sharply peaked at certain energy with a $\delta$-like shape. For clarity, we shaded this peak that dominates the EPC. It corresponds to the SH$^{*}$ (shear horizontal optical) in-plane C-C stretching mode. As the tensile strain increases, on the one hand, the shaded peak moves towards lower energy, reflecting the softening of this particular optical mode\cite{marianetti2010failure}; on the other hand, the shaded peak value is intensified. From Eq. (4), we see that both the red shift (decreasing $\omega$) and the increase of peak intensity (increasing $\alpha^{2}F(\omega)$) will increase $\lambda$.

The spectral features in Fig. 2(a), (c) and (e) suggest that the strain induced phonon softening plays a key role in the strain enhanced EPC. To further quantify the $\lambda-\varepsilon$ relation, we define a characteristic phonon mode ($\omega_{0}$) by averaging over all phonon modes weighted by the Eliashberg spectral function $\alpha^{2}F(\omega)$,
\begin{equation}
\omega_{0}=\langle\omega^{2}\rangle^{1/2}=\sqrt{\int d\omega\,\omega\alpha^{2}F(\omega)/\int\frac{d\omega\,\alpha^{2}F(\omega)}{\omega}} \label{eq:five}
\end{equation}
i.e., each phonon mode is weighted by its EPC strength, so that the calculated $\omega_{0}$ represents the average phonon mode contribution to $\lambda$. The calculated results (data points) of $\omega_{0}$ as a function of strain are shown in Fig. 2(b), (d) and (f) for the $1.55\times10^{14}$ cm$^{-2}$, $3.10\times10^{14}$ cm$^{-2}$, and $4.65\times10^{14}$ cm$^{-2}$ hole-doped graphene, respectively. Clearly, the $\omega_{0}$ decreases with the increasing $\varepsilon$ and can be fit nicely by $\omega_{0}(\varepsilon)=\omega_{0}(\varepsilon=0)+p\varepsilon+q\varepsilon^{2}$, as mentioned above.

Also plotted in Fig. 2(b), (d) and (f) are $\lambda$ as a function of ${\omega_{0}}^{-2}$, illustrating the scaling relation of $\lambda \sim {\omega_{0}}^{-2}$. Thus, we arrived at the empirical formula of Eq.~(\ref{eq:two}) used to fit the data in Fig. 1(b). The exponent $-2$ in the $\lambda\sim\omega_{0}$ relation is more exotic, and serves as the key to understand the nonlinear enhancement of EPC by strain in graphene. In view of the $\delta$-like shaped Eliashberg function that dominates the EPC (Fig. 2), we can assume that the characteristic phonons can be approximated by the Einstein model, all having the same energy $\omega_{0}$. As a nonpolar centrosymmetric crystal, the only EPC type in graphene is the deformation potential interaction\cite{grimvall1981electron}
\begin{equation}
|g_{\textbf{k}+\textbf{q},\textbf{k}}^{\textbf{q}\nu}|^{2}={|\langle \textbf{k}+\textbf{q}|D_{\textbf{q}\nu}\delta u|\textbf{k}\rangle|}^{2}=\frac{{|\langle \textbf{k}+\textbf{q}|D_{\textbf{q}\nu}|\textbf{k}\rangle|}^2}{2M\omega_{0}} \label{eq:six}
\end{equation}
where $D_{q\nu}$ is the deformation potential operator associated with the phonon mode ($\textbf{q}$, $\nu$) and $\delta u=\sqrt{\frac{1}{2M\omega_{0}}}$  is the zero-point oscillation amplitude of a quantum particle with mass $M$.

Considering that $\lambda$ is defined by the EPC-induced renormalization of electron energy spectrum around the Fermi surface, $\lambda \sim \delta E(\textbf{k})/[E(\emph{k})-E(\textbf{k}_{F})]$, we roughly determine the electronic energy shift around the Fermi surface by applying a simple second-order perturbation:
\begin{equation}
\Delta E_{\textbf{k}} \approx \sum_{\textbf{q}, \nu}\frac{|g_{\textbf{k}+\textbf{q}}^{\textbf{q}\nu}|^{2}}{E_{\textbf{k}+\textbf{q}}^0
-E_{\textbf{k}}^0}\delta(E_{\textbf{k}+\textbf{q}}^0-E_{\textbf{k}}^{0}-\omega_{0})
\propto\frac{1}{{\omega_{0}}^2} \label{eq:seven}
\end{equation}
And again, we arrive at $\lambda \sim {\omega_{0}}^{-2}$. Now, we see this relation has two origins. One ${\omega_{0}}^{-1}$ factor comes from the zero-point oscillation amplitude, i.e. softer phonons inducing larger deformation; the other ${\omega_{0}}^{-1}$ factor comes from the energy denominator in the perturbation theory, i.e. softer phonons inducing stronger mixing between different electronic states around the Fermi surface. The former is reflected by the increased shaded peak value in the Eliashberg spectral function under strain (Fig. 2); the latter corresponds exactly to the $1/\omega$ scaling of $\lambda$($\omega$) (Eq.~(\ref{eq:four})).

\begin{figure}[tbp]
\includegraphics[width=0.4\textwidth]{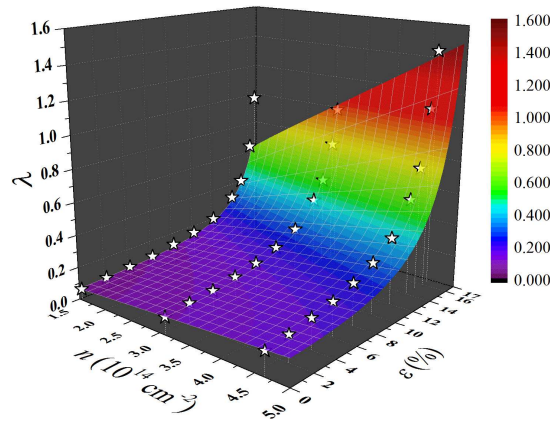}
\caption{\label{fig:fig3} (Color online) 3D plot of $\lambda(n, \varepsilon)$ calculated using Eq. (8), and selected data (stars) calculated from first-prinsiples.}
\end{figure}

Combine Eq.~(\ref{eq:one}) and Eq.~(\ref{eq:two}), we derive an empirical function of $\lambda(n, \varepsilon)$ for the p-type graphene,
\begin{equation}
\lambda(n, \varepsilon)=\frac{\sqrt{n}}{11.93-2.19\sqrt{n}}\cdot\frac{0.5}
{(1-0.007\varepsilon-0.002\varepsilon^{2})^{2}} \label{eq:eight}
\end{equation}
which is applicable to a wide range of doping and strain. Eq.~(\ref{eq:eight}) underlines explicitly the combined effects of doping and strain that greatly enhance the EPC in graphene. Figure 3 shows the 3D plot of $\lambda(n,\varepsilon)$ using Eq.~(\ref{eq:eight}). In comparison, some $\lambda$ values directly calculated from the first-principles (stars) are also shown, and the two agree well.

\begin{figure}[tbp]
\includegraphics[width=0.4\textwidth]{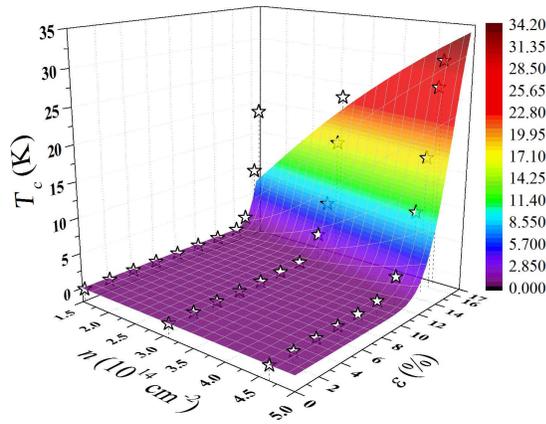}
\caption{\label{fig:fig4}(Color online) 3D plot of $T_{c}(n, \varepsilon)$ calculated using Eq. (9), and selected data (stars) calculated from first-prinsiples.}
\end{figure}

The greatly enhanced $\lambda$ by the combined effects of doping and tensile strain, as shown in Fig. 3, have some very interesting physical implications, including formation of exotic polaronic, charge density wave, and superconducting state. In particular, the possible superconducting state, which may occur at the limit of high doping (high carrier density) and strain (strong e-ph pairing) levels, is very appealing. We have calculated the critical transition temperature for the superconducting state using McMillan-Allen-Dynes formula\cite{allen1975transition}:
\begin{subequations}
\label{eq:whole}
\begin{eqnarray}
T_{c}=\frac{\hbar\omega_{\log}}{1.2k_{B}}\exp[\frac{-1.04(1+\lambda)}{\lambda-\mu^{*}(1+0.62\lambda)}] \label{eq:ninea}
\end{eqnarray}
\begin{equation}
\omega_{\log}=1035.77-38.05\varepsilon-0.70\varepsilon^{2} \label{eq:nineb}
\end{equation}
\end{subequations}
which has been widely used to estimate $T_{c}$ of carbon-based BCS superconductors, such as fullerene\cite{oshiyama1992linear}, carbon nanotubes\cite{iyakutti2006electronic} and intercalated graphite\cite{calandra2005theoretical}. $\omega_{\log}$ is the logarithmically averaged phonon frequency, and we found that $\omega_{\log}$ is almost independent of doping but changes with strain following the empirical relation of Eq.~(\ref{eq:nineb}) \cite{SI}. $\mu^{*}$ is the retarded Coulomb pseudopotential related to the dimensionless parameter of screened Coulomb potential $\mu$ as $\mu^{*}=\mu/[1+\mu \ln(\omega_{e}/\omega_{D})]$, where $\omega_{e}$ and $\omega_{D}$ are the characteristic electron and phonon energy. We have evaluated $\mu^{*}$ for graphene \cite{SI}, which falls in the range of $\sim0.10-0.15$, consistent with the values reported in other carbon-related materials and most $sp$-electron metals.

Using Eqs. (8) and (9) with $\mu^{*}$=0.115, we plot in Fig. 4 the calculated $T_{c}$ as a function of doping ($n$) and strain ($\varepsilon$), superimposed with selected data obtained from first-principles calculations (stars). Most remarkably, at 16.5\% strain, $T_{c}$ reaches as high as 18.6 K, 23.0 K and 30.2 K for the doping level of $1.55\times10^{14}$, $3.10\times10^{14}$ and $4.65\times10^{14}$ cm$^{-2}$, respectively. Such a high $T_{c}$ may appear too surprising at first, but becomes reasonable after one compares with MgB$_{2}$, a well-known BCS superconductor with a theoretically predicted $T_{c}$ $\approx$ 40 K\cite{liu2001beyond} that is in very good agreement with experiment\cite{nagamatsu2001superconductivity}. The characteristic values of $\lambda=1.01$ and $\omega_{\log}=56.2$ meV (453.3 cm$^{-1}$)\cite{liu2001beyond} of MgB$_{2}$ are very comparable to ours for the doped and strained graphene.

Due to the high electron-hole symmetry in graphene about the Dirac point, we expect the electron and hole doped graphene to have similar superconductivity transition under tensile strain. We have calculated $\lambda$ and $T_{c}$ at different tensile strains for both the $4.65\times 10^{14}$ cm$^{-2}$ electron- and hole-doped graphene (See Table S1 \cite{SI}). Very similar trends in $\lambda$ and $T_{c}$ are found, except that the $T_{c}$ in the electron-doped graphene is slightly lower than $T_{c}$ in the hole-doped graphene at the same doping and strain levels.


Our findings uncover yet another fascinating property of graphene with promising implications in graphene-based devices, such as superconducting quantum interference devices and transistors. We reiterate that the superconductivity transition of doped graphene we discover here is triggered by the enhanced EPC under tensile strain, which is fundamentally different from that in metal-decorated graphene\cite{profeta2012phonon}. The enhancement of $\lambda$ in metal decorated graphene arises from additional metal-related electronic states around the Fermi level, which couple strongly in part with the phonon modes of adsorbed metal atoms. In this sense, the superconductivity in metal-decorated graphene is "extrinsic", arising from additional properties of foreign metal atoms and it is similar to their 3D counterparts, such as the intercalated graphite\cite{csanyi2005role, weller2005superconductivity} and MgB$_{2}$\cite{an2001superconductivity}; while the superconductivity in our case is "intrinsic", arising solely from the intrinsic graphene properties modified by doping and strain. Specifically, tensile strain hardly modifies the electronic structure except changing the slope of $\pi$ band, i.e., Fermi velocity, and doping is within the $\pi$ band too. So the electrons scattered by the phonons still consist of $\pi$ electrons of graphene. Apparently, ours is also very different from the chiral superconducting state of the exceedingly high-doped graphene up to the van Hove singularity point \cite{nandkishore2012chiral}.

Finally, it is important to stress that the high $T_{c}$ is achieved within the experimental accessible doping and strain levels. Either chemical doping by adsorption\cite{yokota2011carrier} or electrical doping by gating in a field effect transistor\cite{efetov2010controlling} has shown doping levels above $10^{14}$ cm$^{-2}$ in graphene. On the other hand, as the strongest 2D material in nature, experimentally graphene has been elastically stretched up to $\sim$25\% tensile strain without breaking\cite{lee2008measurement}. It is also interesting to note that a recent theoretical work shows that either electron or hole doping can in fact further strengthen graphene to reach even higher ideal strength under tensile strain\cite{si2012electronic}. Therefore, it is highly reasonable to anticipate experimental realization of high $T_{c}$ superconducting graphene of $\sim30$ K, making graphene a commercially viable superconductor.


\end{document}